\documentclass[aps,prb,preprint,groupedaddress,showpacs]{revtex4}

\usepackage{graphicx}
\usepackage{dcolumn}
\usepackage{bm}
\usepackage{subfigure}

\begin{document}


\title{Subextensive singularity in the 2D $\pm J$ Ising spin glass}


\author{Ronald Fisch}
\email[]{ron@princeton.edu}
\affiliation{382 Willowbrook Dr.\\
North Brunswick, NJ 08902}


\date{\today}

\begin{abstract}
The statistics of low energy states of the 2D Ising spin glass with
+1 and -1 bonds are studied for $L \times L$ square lattices with $L
\le 48$, and $p$ = 0.5, where $p$ is the fraction of negative bonds,
using periodic and/or antiperiodic boundary conditions.  The
behavior of the density of states near the ground state energy is
analyzed as a function of $L$, in order to obtain the low
temperature behavior of the model.  For large finite $L$ there is a
range of $T$ in which the heat capacity is proportional to $T^{5.33
\pm 0.12}$.  The range of $T$ in which this behavior occurs scales
slowly to $T = 0$ as $L$ increases.  Similar results are found for
$p$ = 0.25.  Our results indicate that this model probably obeys the
ordinary hyperscaling relation $d \nu = 2 - \alpha$, even though
$T_c = 0$. The existence of the subextensive behavior is attributed
to long-range correlations between zero-energy domain walls, and
evidence of such correlations is presented.

\end{abstract}


\maketitle


In 1977, Thouless, Anderson and Palmer\cite{TAP77} (TAP) performed a
mean-field theory analysis of the ring diagrams which contribute to
the free energy of the Ising spin glass.\cite{EA75,ICM79}  They
found that, above the critical temperature $T_g$, the contribution
of these ring diagrams was subextensive.  This means that, while the
sum of these diagrams is divergent at $T_g$, their contribution at
any $T > T_g$ can be neglected in the thermodynamic
limit.\cite{ICM79}  Therefore, in this limit, no signature of the
transition is visible in the equilibrium thermodynamic functions for
$T > T_g$.  However, one can still study the critical scaling
behavior of finite systems.

While it is true that hyperscaling is always violated in a
mean-field theory, TAP showed that a spin glass has severe
fluctuations of the order parameter even at the mean-field level.
Later, it was shown by Sompolinsky and Zippelius\cite{SZ81,SZ82}
that the Ising spin glass also violates the fluctuation-dissipation
theorem. Thus one should not be surprised if it turns out that the
spin glass does not obey other relations which work for ordinary
phase transitions.

In this work we analyze data obtained from exact calculations of the
density of low-energy states for finite two-dimensional (2D)
lattices.  The same data have also been used to study the scaling
behavior of domain walls for this model.\cite{Fis06b}  We will
discover that an unusual effect, similar to the violation of
hyperscaling found in mean-field theory, also occurs in 2D. The data
were obtained using a slightly modified version of the computer
program of Vondr\'{a}k,\cite{GLV00,GLV01} which is based on the
Pfaffian method. Our data are completely consistent with the data of
Lukic {\it et al.},\cite{LGMMR04,JLMM06} which were obtained using
the same algorithm.  Our analysis of the heat capacity is more
detailed than theirs, however, and thus we arrive at somewhat
different conclusions.

In two dimensions (2D), the spin-glass phase is not stable at finite
temperature.  Because of this, it is necessary to treat cases with
continuous distributions of energies (CDE) and cases with quantized
distributions of energies (QDE) separately.\cite{BM86,AMMP03}  In
this work we will study the QDE case.

The Hamiltonian of the EA model for Ising spins\cite{EA75} is
\begin{equation}
  H = - \sum_{\langle ij \rangle} J_{ij} \sigma_{i} \sigma_{j}  \,
  ,
\end{equation}
where each spin $\sigma_{i}$ is a dynamical variable which has two
allowed states, +1 and -1.  The $\langle ij \rangle$ indicates a
sum over nearest neighbors on a simple square lattice of size $L
\times L$.  We choose each bond $J_{ij}$ to be an independent
identically distributed quenched random variable, with the
probability distribution
\begin{equation}
  P ( J_{ij} ) = p \delta (J_{ij} + 1)~+~(1 - p) \delta (J_{ij} -
  1)   \, ,
\end{equation}
so that we actually set $J = 1$, as usual.  Thus $p$ is the
concentration of antiferromagnetic bonds, and $( 1 - p )$ is the
concentration of ferromagnetic bonds.  Here we will discuss
primarily the equal mixture case, $p = 0.5$, but results for $p =
0.25$ will also be given.

The ground state (GS) entropy is defined as the natural logarithm
of the number of ground states.  For each sample the GS energy
$E_0$ and GS entropy $S_0$ were calculated for the four
combinations of periodic and antiperiodic toroidal boundary
conditions along each of the two axes of the square lattice.  When
$p = 0.5$, all four of these types of boundary conditions are
statistically equivalent.

Data were obtained for lattices of sizes $L$ = 7, 8, 11, 12, 15, 16,
21, 24, 29, 32, 41 and 48.  For each $L$, 500 different random sets
of bonds were studied, for each of the four boundary conditions.
Thus, combining the data for the different boundary conditions, we
have 2000 values of $E_0$ and $S_0$ for each $L$.

\begin{figure}
\includegraphics[width=3.4in]{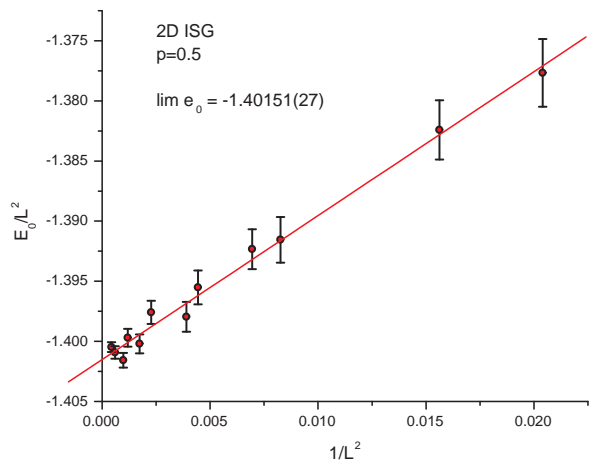}
\caption{\label{Fig.1} (color online) Finite-size scaling of
$E_0 / L^2$ vs. $1/L^2$.}
\end{figure}

With the boundary conditions we are using, for which there is no
well-defined surface, the value of $E_0$ averaged over samples of
the random bonds, is expected to obey
\begin{equation}
  E_0 / L^2 ~=~ e_0~+~a_e / L^2   \,
\end{equation}
to lowest order in $L$.  Fig.~1 shows that this works well, and that
the value of $e_0$ obtained from our data is $e_0 = -1.40151 \pm
0.00027$.  In principle, higher order corrections exist,\cite{CHK04}
but they are not necessary at the level of precision of our data.
This agrees with the result found by Lukic {\it et
al.}\cite{LGMMR04} All statistical error estimates shown in this
work represent one standard deviation. The best estimate of $e_0$ is
still the one of Palmer and Adler,\cite{PA99} which uses a method
for which one can go to much larger $L$, because the entropy is not
calculated.

\begin{figure}
\includegraphics[width=3.4in]{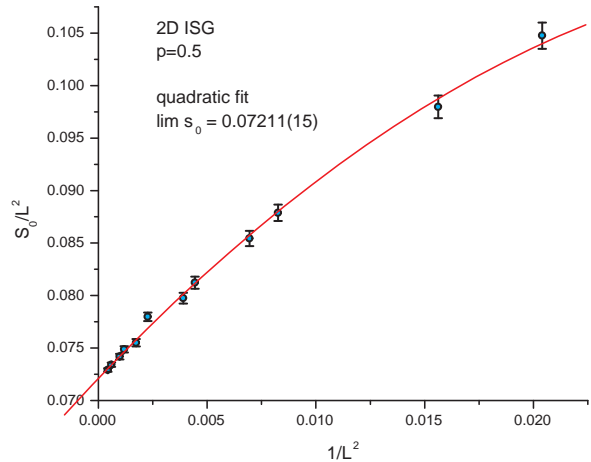}
\caption{\label{Fig.2} (color online) Finite-size scaling of
$S_0 / L^2$ vs. $1/L^2$.}
\end{figure}

The finite-size scaling behavior of $S_0$ is slightly more complex.
Lukic {\it et al.}\cite{LGMMR04} used a single correction-to-scaling
term, with an exponent $-(2 + \Theta_S)$.  From a fundamental
viewpoint,\cite{BKM03} however, when $\Theta_S$ is positive the
natural form to use when adding another fitting parameter is
\begin{equation}
  S_0 / L^2 ~=~ s_0~+~a_s / L^2~+~b_s / L^4   \, .
\end{equation}
In Fig.~2 we see that this form works well, and gives a value of
$s_0 = 0.07211 \pm 0.00015$.  This value is slightly higher than the
one quoted by Lukic {\it et al.}, but the difference comes primarily
from the different form of the fitting function rather than from
differences in the data.  By comparing with the work of Bouchaud,
Krzakala and Martin,\cite{BKM03} one sees that Lukic {\it et al.}
have made a sign error, and that their fit actually uses a negative
value for $\Theta_S$, which is incorrect.\cite{SK93,SK94}

While our values of the energy and entropy of the GS of finite $L
\times L$ lattices for $p = 0.5$ are generally consistent with those
of other workers, our results for $L = 32$ differ substantially with
those reported by Blackman and Poulter.\cite{BP91}  (See Figs. 7 and
8 of their paper.)  The origin of this discrepancy is unclear, but
it appears to be too large to be explained by the different boundary
conditions used by them.  Their numbers of samples computed are
rather small, and it may be that they have simply underestimated
their statistical errors.  However, their algorithm, unlike the one
used here, does not use exact integer arithmetic to calculate the
partition function.  Therefore, it is likely that they have a
problem with roundoff errors.  In a strongly correlated system such
as the one we are studying, substantial roundoff errors can result
in distributions which are too narrow.

\begin{figure}
\includegraphics[width=3.4in]{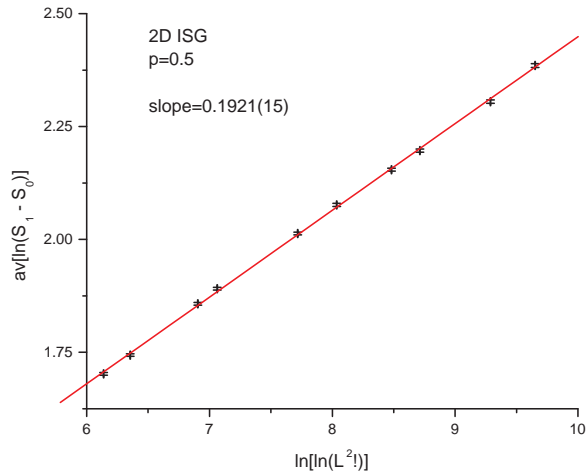}
\caption{\label{Fig.3} (color online) Scaling of $S_1 - S_0$ with
$L$.}
\end{figure}

In order to obtain information about the low temperature behavior,
it is useful to study the scaling with $L$ of $S_1 - S_0$, which
is the logarithm of the ratio of the degeneracies of the lowest
excited state and the GS.\cite{WS88,SK93,SK94,LGMMR04}  We found that
\begin{equation}
  {\rm av}(\ln(S_1 - S_0)) ~=~ \rho \ln(\ln((L^2)!)) ~+~ 0.528 \pm
  0.011  \, ,
\end{equation}
with
\begin{equation}
  \rho ~=~ 0.1921 \pm 0.0015
\end{equation}
gives an excellent fit for $L > 10$, as shown in Fig.~3.  av() is a
configuration average over random samples.  The points for $L =$ 7
and 8 (not shown in the figure) are below the fitted line, due to
corrections to scaling at small $L$.

The choice $\ln(\ln((L^2)!))$ may appear arbitrary to the reader,
but it was suggested by the behavior of the fully frustrated 2D
Ising model.\cite{SK93,SK94} In principle, if one could go to very
large values of $L$, one could obtain $\rho$ by plotting the data
against $2 \ln (L)$.  From Stirling's approximation one sees
immediately that the difference between using $2 \ln (L)$ and
$\ln(\ln((L^2)!))$ is a logarithmic correction to scaling.  This
logarithmic correction appears to be present in the data, however,
and a much better fit is obtained if one does things as shown here.

If one uses $2 \ln (L - 3)$ an excellent fit over the range of the
data is obtained.  However, this seems completely artificial to the
author.  In any case the value of $\rho = 0.1948 \pm 0.0008$ which
one finds from this form is close to the one shown in Fig.~3.  (The
reason why the statistical error in this number is so small is that
no contribution from the uncertainty in the fitting parameter $``3"$
is included.)

\begin{figure}
\includegraphics[width=3.4in]{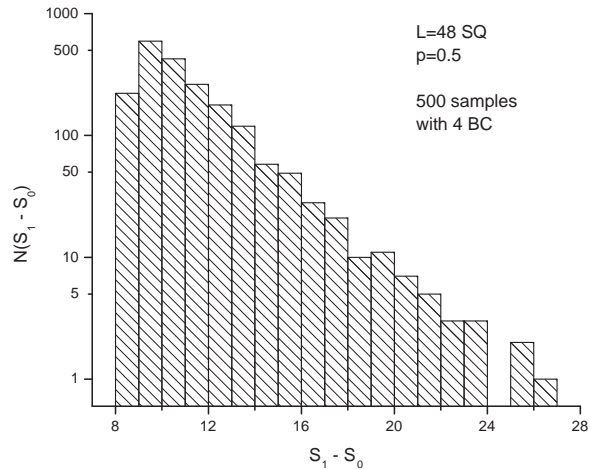}
\caption{\label{Fig.4} Histogram of the distribution of $S_1 - S_0$
for $L=48$.}
\end{figure}

The reason for taking the configuration average of $\ln(S_1 - S_0)$
rather than taking the logarithm of av$(S_1 - S_0)$ is that in this
way we find the most probable value.\cite{DH81,MC02}  The
probability distributions for $S_1 - S_0$ are highly skewed, and the
most probable value scales differently with $L$ than av$(S_1 - S_0)$
does.  To illustrate this point, in Fig.~4 we show a histogram of
the distribution of $S_1 - S_0$ for the $L=48$ lattices.  If one
plots the data using av$(S_1 - S_0)$, one finds an apparent value
for $\rho$ of 0.233(3).  Using the median value gives 0.222(3).  It
is the typical or most probable value which is the experimentally
observable quantity, as established by Edwards and
Anderson\cite{EA75} for the spin glass.

From this analysis, we obtain the typical value of $S_1 - S_0$ to be
\begin{equation}
  S_1 - S_0 ~=~ f(L) ~\approx~ A [\ln((L^2)!)]^{\rho}  \, ,
\end{equation}
with $A = 1.696 \pm 0.019 $, or, using Stirling's approximation,
\begin{equation}
  f(L) ~\approx~ A [L^2 (2 \ln(L) - 1)]^{\rho}  \, .
\end{equation}
It follows immediately that the scaling of $S_1 - S_0$ with $L$ is
approximately a power law, with an exponent close to 0.4, times
$\ln(L)$.  This variation with $L$ is much more rapid than the
hypothesis of Wang and Swendsen,\cite{WS88} who argued for a
dependence like $4 \ln(L)$.  To this extent, it agrees with the
claims of J\"{o}rg {\it et al.}\cite{JLMM06}

To obtain the actual behavior of the low temperature specific heat,
we must carry the analysis further.  The heat capacity of a sample
of size $L \times L$ at temperature $T$ is given by
\begin{equation}
  C(L,T) ~=~ \langle ( E(L) - \langle E(L) \rangle )^2 \rangle /
 T^{2} \, ,
\end{equation}
where the angle brackets indicate a thermal average, and we are
using units in which Boltzmann's constant is 1.

Writing the partition function of a finite sample with periodic
boundary conditions explicitly gives
\begin{equation}
  Z(T) ~=~ \sum_{n = 0}^{| E_0 | / 2} \exp(S_n - S_0~-~4 n / T) \,
  .
\end{equation}
The heat capacity is then
\begin{equation}
  C(L,T) ~=~ (T^2 Z)^{-1} \sum_{n = 0}^{| E_0 | / 2} 16(n - n_*)^2
 \exp(S_n - S_0~-~4 n / T) \, ,
\end{equation}
where $n_*$ is the value of $n$ for which the argument of the
exponential has its maximum for a given sample at temperature $T$.

\begin{figure}
\includegraphics[width=3.4in]{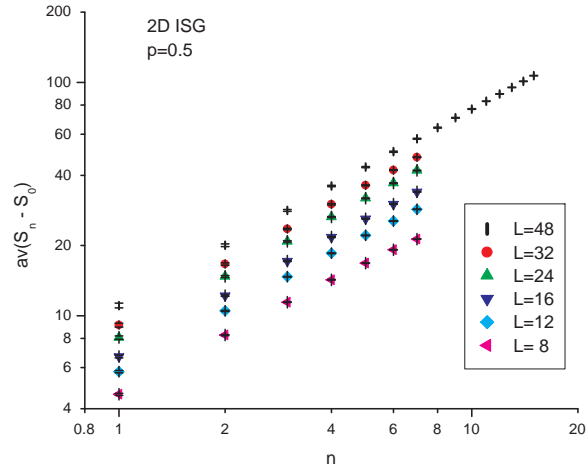}
\caption{\label{Fig.5} (color online) Scaling of av$(S_n - S_0)$
with $L$, for small values of $n$.  The axes are scaled
logarithmically.}
\end{figure}

\begin{figure}
\includegraphics[width=3.4in]{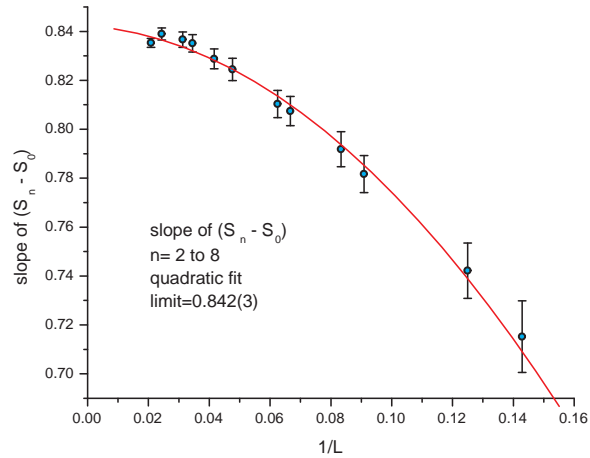}
\caption{\label{Fig.6} (color online) Slope of av$(S_n - S_0)$ vs.
$1/L$, for $n$ = 2 to 8.}
\end{figure}

The average values of $S_n - S_0$ for small values of $n$ are shown
in Fig.~5, over our full range of $L$.  The slope defined by these
points, omitting the $n = 1$ points, is plotted versus $1/L$ in
Fig.~6. The limiting value of this slope for large $L$ obtained from
this plot is found to be
\begin{equation}
  \psi ~=~ 0.842 \pm 0.003 \, .
\end{equation}
This means that for $n \ll L^2$
\begin{equation}
  S_n - S_0 ~\approx~ f(L) n^{\psi} \, ,
\end{equation}
and implies
\begin{equation}
  (n_*(L,T))^{1 - \psi} ~\approx~ \psi f(L) T / 4 \, .
\end{equation}
This can only be valid, however, if $0~<~n_*~\ll~L^2$.  If we take
the limit $T \to 0$, holding $L$ fixed, then $n_* \to 0$.  Thus
the limiting low temperature behavior of $C(L,T)$, for any fixed
$L$ is proportional to $\exp( -4 / T)$, as it must be.  We expect
to see this behavior when $T < T_1$, where
\begin{equation}
  T_1 ( L ) ~=~ 4 [L^2 ( 2 \ln(L) - 1 )]^{- \rho} / (\psi A)  \,
\end{equation}
is the temperature where $n_* = 1$.  We have found a positive value
for $\rho$, which means that $T_1 \to 0$ as $L \to \infty$.

The reason for omitting the $n = 1$ points shown in Fig.~5 from the
fits is that they all lie well below the straight lines.  The
quantity $S_1 - S_0$ does not behave in the same way that the other
$S_{n+1} - S_{n}$ do.  The author understands this effect by analogy
with the well-known behavior of random matrices.  The gap at the
band edge is special, because it only feels level repulsion from one
side.

Substituting our expressions for $n_*$ and $S_n - S_0$ into
Eqn.~(10) gives
\begin{equation}
  C(L,T) ~=~ 16 T^{-2}
   {\sum_{n = 0}^{| E_0 | / 2}(n - (\psi A T / 4 )^{1/(1-\psi)}
  (L^2 (2 \ln(L) - 1))^{\rho/(1 - \psi)})^2 \exp(g(n,L)) \over
   \sum_{n = 0}^{| E_0 | / 2} \exp(g(n,L))}
    \, ,
\end{equation}
where
\begin{equation}
  g(n,L) ~=~ n^{\psi}[ A (L^2 (2 \ln(L) - 1))^{\rho} ~-~
  4 n^{1 - \psi} / T ]  \, .
\end{equation}

When we try to take the limit $L \to \infty$ holding $T$ fixed, we
get a surprise.  The exponent $\rho/(1 - \psi)$ is $1.216 \pm
0.033$. Because this exponent is greater than than 1, the power-law
behavior described by the exponent of Eqn.~(12) is only valid for
$T < T_x$, where $T_x$ must go to zero as $L$ increases. $n_*$ {\it
cannot} become larger than $L^2$! This condition requires that,
when $L \to \infty$, $T_x$ must also go to zero at least as fast as
\begin{equation}
  T_x (L) ~\sim~ 4 L^{-2 (\rho + \psi - 1)} ( 2 \ln(L) - 1 )^{-
  \rho} / (\psi A)   \, .
\end{equation}
Although we do not have data to show that $T_x$ actually behaves
precisely in this way, it is at least plausible that $T_x$ goes to
zero more slowly than $T_1$ as $L$ increases, since $\psi < 1$.

What this means is that the singularity we are studying is
subextensive, just as the thermal singularity above $T_g$ is in the
TAP mean-field theory.\cite{TAP77,ICM79}  It also means that for $L$
large, but finite, we expect there exists a temperature regime $T_1
\ll T \ll T_x$ in which the scaling behavior controlled by this
singularity is observable.

$\rho$ controls the thermal behavior in the temperature range $0 < T
< T_1$, and $\psi$ controls the behavior in the range $T_1 < T <
T_x$.  Therefore, these exponents are independent.  A simple scaling
relation between exponents defined in different ranges of T which
have independent behaviors is impossible.  This statement is not in
contradiction with the fact that the value of $T_x$ clearly depends
on both $\rho$ and $\psi$.  The entire procedure used here is quite
similar to the theory of nested boundary layers.\cite{BO78}

Since $\langle E (L) \rangle$ is essentially $4 n_*(L,T)$, the heat
capacity for $T_1 < T < T_x$ is easily seen to be proportional to
$T^{\psi / (1 - \psi)}$, which is $T^{5.33 \pm 0.12}$.  Because $T_x
\to 0$ as $L \to \infty$, this behavior disappears in the
thermodynamic limit. The exponent $2 (\rho + \psi - 1)$ is $0.068
\pm 0.009$.  This is small, so $T_x$ is going to zero quite slowly.
Thus the power-law behavior of $C (L,T)$ should be visible for
macroscopic values of $L$.  Note that this effect is not caused by
our choice of logarithmic averaging of $S_1 - S_0$, since the use of
simple averaging would give a larger value for $\rho$.

Although our statistical errors are small, the estimate of $\rho$
depends on our choice of the finite-size scaling fitting function.
Notice that the estimate of the scaling exponent $\alpha$ for the
$T$ dependence of $C (L,T)$ depends only on $\psi$, and is
independent of $\rho$. Therefore, our estimate
\begin{equation}
 \alpha ~=~ -5.33 \pm 0.12  \,
\end{equation}
is independent of whether $T_x \to 0$ as $L \to \infty$.

All of the calculations for $p=0.5$ described above were repeated
for $p=0.25$.  Using the same procedures as discussed above, we find
for $p=0.25$ the exponents $\rho = 0.1874 \pm 0.0019$ and $\psi =
0.8527 \pm 0.0017$. Therefore we obtain $2 (\rho + \psi - 1) = 0.080
\pm 0.007$ and $\alpha = -5.79 \pm 0.08$.  These results are quite
consistent with universality of the critical exponents, since the
quoted statistical errors do not include any allowance for errors in
the assumed scaling forms.

Recently, J\"{o}rg {\it et al.}\cite{JLMM06} have claimed that a
power-law behavior of $C (L,T)$ is evidence that the QDE is in the
same universality class as the CDE.  However, they have not
calculated $\alpha$ directly.  They have calculated the correlation
length exponent $\nu \approx 3.5$, and assumed that $\alpha$ could
be obtained via the modified hyperscaling relation of Baker and
Bonner.\cite{BB75}  The fact that our value of $\alpha$ is not close
to $-7$ shows that this relation is not obeyed. Our value seems to
indicate that the ordinary hyperscaling relation, $d \nu = 2 -
\alpha$, is obeyed.  $\alpha$ has never been calculated directly for
the CDE, so we cannot say whether the values of $\alpha$ are the
same for the QDE and the CDE.

\begin{figure}
\includegraphics[width=3.4in]{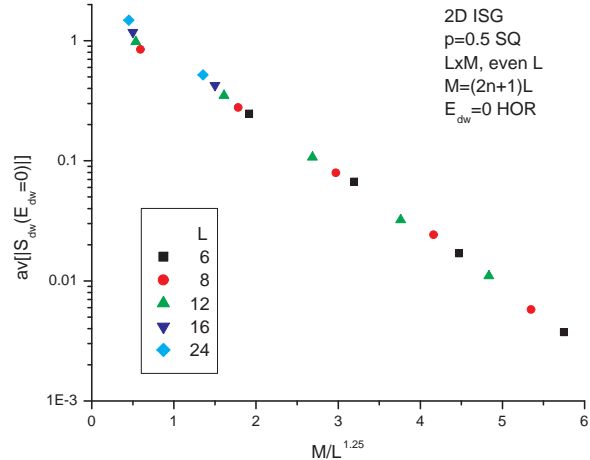}
\caption{\label{Fig.7} (color online) Scaling of average entropy of
zero-energy domain walls for lattices of size $L \times M$ vs. $M /
L^{1.25}$. These domain walls run across the lattice in the short
direction, which has length $L$.}
\end{figure}

Finally, we discuss the origin of the subextensive singularity. Such
behavior in a 2D model probably requires the existence of some kind
of long-range interactions.  Such interactions are not present
explicitly in our Hamiltonian, Eqn.~(1), but they may arise
spontaneously. Since domain walls are extended objects, it would not
be very surprising for interactions between domain walls to have
long range, especially at $T = 0$.

Using the same computer program which was used here to obtain the
heat capacity and additional procedures described in a recent
publication,\cite{Fis06b} we have calculated the average domain-wall
entropy for zero-energy domain walls on lattices of size $L \times
M$, where $L \le M$.  Remarkably, the average domain-wall entropy
for the zero-energy domain walls which run across the lattice in the
short ($L$) direction scales to zero exponentially in the variable
$M / L^{1.25}$. This is shown in Fig.~7.  The exponent 1.25 is
suggestive of the relation recently proposed by Amoruso, Hartmann,
Hastings and Moore,\cite{AHHM06} which gives a value of 1.25 for the
fractal dimension of domain walls for this model.  From the data
displayed here we can say that this exponent must be $1.25 \pm
0.05$.  Because their entropy scales to zero so rapidly, these
zero-energy domain walls must be highly correlated.

This effect is strong evidence for long-range interactions between
the zero-energy domain walls. It does not occur for domain walls of
other energies.  Amoruso {\it et al.} do not explicitly specify that
the behavior of the zero-energy domain walls should be special.
However, this was suggested by the work of Wang, Harrington and
Preskill.\cite{WHP03}  This domain-wall entropy calculation will be
described more fully in a subsequent publication.\cite{Fis07}

In this work we have calculated in detail the low temperature
thermal behavior of the 2D Ising spin glass with an equal mixture of
$+1$ and $-1$ bonds.  We have found that this behavior bears a
strong qualitative resemblance to the behavior found in the TAP
mean-field-theory analysis.  For finite $L$ there is a range of $T$
for which $C (L,T)$ is proportional to $T^{5.33}$.  However, this
behavior disappears slowly as $L \to \infty$.  This subextensive
behavior is attributed to correlations between zero-energy domain
walls.

\begin{acknowledgments}
The author thanks Jan Vondr\'{a}k for providing his computer code,
and for help in learning how to use it.  He is grateful to David
Huse and Alex Hartmann for stimulating conversations, to Mike Moore
for finding an error in the manuscript, and to Princeton University
for providing use of facilities.

\end{acknowledgments}



\end{document}